\documentclass[a4paper,11pt]{article}
\usepackage{graphicx}
\usepackage{dcolumn}
\newcommand{\beq}{\begin{equation}}
\newcommand{\eeq}{\end{equation}}

\title{On the classical limit of Quantum Mechanics}
\author{M. Baldo}
\date{ }
\begin{document}
\maketitle
\vspace {-1 cm}
\begin{center}
\textit{Istituto Nazionale di Fisica Nucleare}
\par \textit{via Santa Sofia 64, 95123 Catania, Italy} 
\end{center}
\vskip 0.5 cm
\noindent Abstract.
\par 
One of the main unsolved question in Quantum Mechanics (QM) is its compatibility with classical mechanics. The laws of
QM, which describe the microscopic world, must merge into the classical ones for large enough physical systems,
but it is still unknown at which point, if any, the transition occurs and if the transition is smooth or sudden as the size increases.
Furthermore, a strict extension of QM to the macroscopic world leads to well known 'paradoxes', which is not straightforward
to solve. This question is tightly connected with the measurement problem, since any measurement apparatus must give a response which can be described at classical level. Many experiments have been performed to answer to this question. They mainly try to find to which extent the number of particles can be increased in order to observe clear evidence of violation of standard QM.
In this paper we argue that the macroscopicity parameter is not the number of particles but the number of independent degrees of freedom, as proposed in a recent model for the completion of QM. This introduces a sort of change of paradigm. To support this claim a representative set of well known experiments are analyzed from this point of view. This brings to a new interpretation of the experiments. A general scheme for the quantum-classical transition is discussed in some details.
\section{Introduction.\label{sec:int}}
One of the most basic question on Quantum Mechanics (QM), which has not yet received a definite answer, is the connection
of QM with classical physics. The laws of QM are quite different from the classical ones, but it is certainly true that they are both correct. Therefore it is legitimate to ask how they can be compatible, which is equivalent to ask in which way QM reduces to the classical physics that describes so accurately the world of macroscopic objects. It is true that in some cases,
like in superfluids or in superconductors, QM manifests at macroscopic scale, but in general the laws of QM seem to have no relevance in the macroscopic world. There can be different general schemes to address this basic question,
but they are all still debatable. The transition to classical physics is intimately connected with the "measurement problem"
in QM, since the response of a proper detector must be described in classical terms.\par 
It is possible that the laws of QM are actually valid to all scales, but their manifestations
become more and more subtle as the macroscopic scale is approached, until they become undetectable. This is the case of the short wave length limit, where the values of the actions present in a system are much larger than the Plank' constant $\hbar$. However this assumption is
difficult to maintain. If the superposition principle is assumed to be valid at any scale some paradoxes arise, noticeably the so called "Schrodinger cat paradox" \cite{sch,cat}. Furthermore, in a genuine selective measurement the superposition principle is
abruptly violated due to the process of wave function reduction.\par 
Another scheme that has been proposed is the "continuous spontaneous collapse" (CSL) model \cite{CSL_RMP}. In this case an additional QM process is introduced, where the wave function of a generic system undergoes a spontaneous reduction in coordinate space.
The process is assumed to be random, with a probability that increases with the size of the system, in particular in the measurement processes, where the response of the apparatus must reach the macroscopic scale. However the CSL model cannot explain the so called null experiments, i.e. reduction without interaction, as well as spontaneous emission. Furthermore CSL deals only with reduction of the wave function in coordinate space and it does not predict reduction in other 
representations.\par 
The environment has been invoked as the main actor that produces the wave function reduction, and possibly the disappearance
of QM in favor of classical mechanics. In fact it is not difficult to show that a random external perturbation can destroy the coherent superposition of states, which is the first step toward the transition to classical mechanics and the description of the measurement process. However the decisive next step is the selection of a single component to complete the reduction 
process and it is hard to invoke the environment for this selection.  In fact no formalism within this scheme has been 
devised which is able to fully describe the reduction process, especially if one demands the fulfillment of the Born' rule.
Furthermore it is unclear how the action of the environment is able in a general context, other than measurement, to produce the transition to the laws of classical mechanics. An underlying difficulty is the ambiguity in the choice of the representation, since different representations can give different outcomes and a rule for adopting the "correct" representation is needed.  \par 
To solve the problem of how the basis of the outcomes is selected in quantum measurement, the so called "Darwinism" model has been proposed \cite{Darwin}.
Schematically, this elaborate modeling of the measurement process introduces, besides the environment decoherence process, a further step, based on the "redundancy" of the possible observations of the system, due to the fragmentation of the environment. It is this redundancy that shapes the objectivity and possible classical behavior of the observed system. The model seems to blur the boundary between the perception of reality and reality itself. However, the model does not provide a mechanism, i.e. a physical process, that is able to select the unique output in a single measurement.\par 
Several experiments have been performed in order to clarify the quantum-classical transition. The purpose of most of them  
was to check to what extent quantization and the quantum laws are still valid for macroscopic systems. To the author knowledge no one of these experiments has found evidence of the failure of QM for large enough system. In particular the state quantization, the superposition principle and the quantum tunneling were found to be valid even for systems with  macroscopically large number of particles. In order to be able to develop further the discussion on the quantum-classical transition, in the next section a representative selection of these experiments will be schematically reviewed. 
\section{Experiments.\label{sec:exp}}
In this section we consider a set of experiments, considered as representative of the different methods employed in this basic context, without the pretension to be exhaustive.\par 
A series of experiments \cite{C60,matter_wave1,matter_wave2,25kDa,nano_review,levitated_nanopart} were devised to test
if large objects, that eventually could be considered macroscopic, follow the basic laws of QM, in particular the
superposition principle. Matter wave interference has been observed for large molecules \cite{C60,matter_wave1,matter_wave2}, up to sizes \cite{25kDa} that, to a certain extent, can be classified as "macroscopic". 
More recently \cite{nanopart} interference experiments has been extended to nanoparticles containing more than 7000 atoms and masses greater than 170,000 Da. 
This implies that the superposition principle is still valid for these massive objects, which therefore preserve their wave nature. Of course, due to the short wave lengths, the interference fringes become quite narrow and dense, but they are still detectable. To increase further the size of the objects appears challenging.  
\par 
We can schematize these experiments by referring to the basic two slits experiment, where an incoming molecular beam is split into two
coherent branches, which follow two different paths and then again joined to interfere. Along the way to the detection the
quantum state of the molecule can be represented by the density matrix
\beq
\rho \,=\, \rho_{in}\,\, \rho_{CM}
\label{eq:den}
\eeq 
\noindent which is the product of the density matrix for the internal degrees of freedom, and the density matrix for the center of mass coordinate. The latter corresponds to a pure state, which is the superposition of the two wave packets pertaining to the two paths
\begin{eqnarray}
\rho_{CM} &\,=\,& | w > < w | \\ \nonumber
\ &\ & \\ \nonumber
w(\mathbf{R}) &\,=\,& \frac{1}{\sqrt{2}} \big( w_1(\mathbf{R}) \,+\, w_2(\mathbf{R}) \big)
\label{eq:sup}
\end{eqnarray}
\par\noindent The internal density matrix is the same for the two branches, since they are equal at the position of the splitting. Any interaction that modifies the internal density matrix in one of the branch would destroy interference.  
 The probability $P(\mathbf{R})$ to detect the molecule at the point $\mathbf{R}$ is obtained by tracing the projection operator $J_R \,=\, | \mathbf{R} > < \mathbf{R} | $
\beq
 P(\mathbf{R}) \,=\, Tr(\rho\, J_R) \,=\, \frac{1}{2} | w_1(\mathbf{R}) \,+\, w_2(\mathbf{R}) |^2  
\label{eq:prob}
\eeq
\noindent where we have taken $ Tr(\rho_{in}) = 1 $. The interference pattern, inherent to Eq. (\ref{eq:prob}), is therefore the same as for a single particle. It has to be noticed that the molecule or the nanoparticle can be in any mixed state, e.g. it can be thermally excited. This is what is indeed observed, in particular in ref. \cite{C60}.\par 
Another related question about quantum superposition is if it can be maintained at large distance, i.e. to what extent the
distance in the superposition between two localized states of the same quantum object can be stretched. This is equivalent to ask how large can be the size of a single object wave packet. The experiment of ref. \cite{large} was able to show
that the superposition of two localized components of an atomic state can be kept up to a distance of 53 cm,
for a time long as much as 1 sec. This is akin to the observation of neutron interferometry with branches at macroscopic distances \cite{neut_int}. However, in this case the system is a cloud of cold atoms at very low temperature. The cloud is of bosonic atoms and therefore is in a superfluid state, with a number of atoms estimated to be around 10$^{5}$, which could be considered as "macroscopic". \par 
According to the theory of superfluidty \cite{NoPi}, near zero temperature and slow enough dynamics, superfluid matter can be described by a single "macroscopic wave function", the so called superfluid amplitude. The latter is just what is usually indicated as "matter wave". There is a very large experimental body of evidence that supports this view. In particular, observations of interference phenomena with this matter wave are quite numerous and experiments on this subject has been developed and implemented along the years\cite{first_interf,first_interferometry,two_BEC,diff_interf,interferometer_BEC,mic_grav}. In the experiment of ref. \cite{large} the atomic cloud can be viewed as a single wave packet of matter wave with two components that are displaced at a distance up to 53 cm. It is worth stressing that the matter wave is described by a function of one single 3D coordinate, as any other wave. The observation of interference fringes in a special version of Mach–Zehnder interferometer is a clear evidence that the description in terms of a single wave is the correct one and that therefore the superposition between the two branch of the interferometer is indeed kept.\par 
The superposition principle for macroscopic systems has been tested also in the experiment of ref. \cite{SQUID1}, 
where a SQUID (Superconducting Quantum Interference Device) was used to implement a "quantum electromagnetic circuit".
In such a device a superconducting ring interrupted by a Josephson junction is placed under a constant magnetic field
perpendicular to the ring. The system is characterized by the capacitance of the Josephson junction and the inductance of the ring and therefore form an effective LC electrical circuit. In short, the total energy is the sum of the magnetic energy and of the "Josephson energy", which are competing between each other. It turns out that, with a suitable choice of the physical parameters, the total potential of the device as a function of the flux across the ring has a double well shape. The two potential minima correspond to two different quantum numbers (0 and 1) for the physical quantity called "fluxoid". They correspond to the two possible direction of rotation of the macroscopic Josephson current. The system 
can choose as stable stationary state one of the two minima. Upon quantization one gets two possible quantum ground states,
with negligible tunnel probability between them. In particular, if the system is initially in the ground state of the higher minimum, it will remain indefinitely in that state. However, the system can be excited to an upper level which is still below the potential barrier between the two minima. In this case the tunnel probability is substantial, and the system will transit to the lower potential well. During the transition the system will be in the superposition of two levels, each one belonging to the two different potential pockets. It is this superposition that the experiment is able to detect. All that is to stress that the superposition involves two opposite direction of rotation of the current, i.e. two different macroscopic configurations.     
\par 
This experiment leans on a series of previous experiments \cite{first_circuit, second_circuit,delta_circuit,review1,review2,review3}, which have shown that indeed electromagnetic circuits, which include a superconductive part, are indeed quantized. This is a basic and highly not obvious result, which paved the way to the physics of qbits.  \par 
Another road to the test of quantization and the superposition principle is the use of mechanical systems, where mechanical oscillations of macroscopic object are found to be indeed quantized and are prepared in a superposition of two eigenstates.  
In ref. \cite{mac_phonon_mode} localized macroscopic vibrations are shown to exhibit quantum features by the tomography on the Wigner function for a single mode and for the superposition of two modes. It is estimated that about 10$^{11}$ particles are involved in the oscillations. However, all the particles undergo the same collective motion, which is a standing phonon state.
The macroscopic oscillation can be viewed as a quantum condensate in a given phonon state, which then corresponds to a configuration of the phonon field with a well defined "macroscopic wave function"  \cite{mac_phonon_mode}. Another experiment \cite{drums1,drums2} demonstrates the validity of the quantization and the superposition principle even at macroscopic level.       
In this case two small, but still macroscopic, vibrating drums are weakly coupled inductively. The astonishing peculiarity
of the experiment is the possibility to measure, with sufficient precision, the position of the drums as they are vibrating.
The two drums has two different proper frequencies and are excited by a radiation flash of frequency intermediate between the two. It is observed that both drums are vibrating with amplitudes that are in quadrature among each other, e.g. they 
reach a maximum at the same time, but with opposite velocities. Also in this case the vibrations are macroscopic and therefore involve a macroscopic occupancy of a definite phononic quantum state. The coupling between the two drums can be schematized as a weak coupling between two quantum levels. As it is well known the weak coupling generates two new eigenstates
which are the symmetric and the anti symmetric linear superposition of the original unperturbed states. If the system is excited to a state which is symmetric linear combination of the two eigenstates, the evolution will be an oscillation between the two eigenstates and the two probabilities that the system is in each one of the original levels are indeed in quadrature,
as it can be verified by a direct calculation. The experimental observation is therefore a check of the quantum nature of the drums vibration, as well as that the two drums are entangled in a common oscillation that involves the superposition principle. 
\par 
One of the most peculiar phenomenon in QM is the tunneling process, and it is therefore natural to ask if its manifestation is present for macroscopic systems, or at least up to which size it can be present. Several experiments 
\cite{macTun1,macTun2,macTun3,macTun4,macTun5,macTun6,macTun7,macTun8,macTunSpin} have been device to answer this question. Most of them consider the tunneling between two voltage or current stationary states of a Josephson junction. They all observe the tunneling process which involves a change in the macroscopic current of the junction. A different system has been considered in ref. \cite{macTunSpin}, where tunneling between the magnetic state of a bulk magnetic material has been studied.  
\section{Discussion.\label{sec:disc}}
From all the experiments mentioned in the previous section one gets the conclusion that the quantization and the superposition principle are still present in physical systems with a number of particles that can be considered as "macroscopic". 
One then could conclude that the transition, if any, from QM to classical mechanics must occur for an even higher number of particles. However, this conclusion is valid under the assumption that the number of particles is the relevant parameter that determines the possible deviation from QM and the transition to classical mechanics, which can be abrupt or smooth at increasing size. An alternative assumption is that the relevant parameter is not the number of particles but rather the number of "active degrees of freedom". The ground state of a system, if not degenerate, is necessarily a pure quantum state, with no
relation to classical physics. Therefore classical behavior, including its center of mass motion, can arise only from the excitation of its internal degrees of freedom. 
In simple terms by "degrees of freedom" we mean one of the variables that appears in the description of the excitation, besides the center of mass. To specify the wave function of the excitation one needs a well defined number of variables, which is independent of the representation. As an example a system of two spinless particles has 6 degrees of freedom in ordinary 3D space. The number of degrees of freedom, so defined, can be much smaller than the number of particles. Consider for instance a rigid body in its ground state. As already discussed in previous section, the total wave function factorizes in the fixed internal wave function and the center of mass wave function. The number of degrees of freedom is 3, i.e. the position of the center of mass. A more general definition of degrees of freedom refers to the parameters that are
needed to describe the evolution of the system. If the wave function is the superposition of different components,
the number of degrees of freedom will be the number of variables that are needed to fix each component (eventually the same number for all components). As an example let us consider a schematic view of a scintillator that measures the positions of particles
within a plane. Once activated by a particle hitting, the evolution of the system can be described by the number of
excited atoms or molecules of the scintillator  along the trace of the particle. This number will be the number of degrees of freedom, which evolves in time. 
\par 
In the experiments mentioned at the beginning of the previous section with large molecules or nanoparticles the number of
degrees of freedom are the ones of the center of mass, which is the same as a single particle. According to the new criterion
in this case no reduction can take place, as observed, at whatever size of the system, if the internal wave function is not modified during the experiment. Similar considerations apply to the experiment of ref. \cite{large} on the superposition principle at large distance. However in this case the number of degrees of freedom is dictated by the appearance of the macroscopic wave function, which contains a single coordinate, and therefore it is the same as for a single particle.
\par 
In the experiments that use a SQUID as the main device, the macroscopic quantity that displays quantum superposition is the superconducting current, which can be in a superposition of two currents in opposite directions, each one involving a macroscopic number of electrons. Also in this case one can show that the number degrees of freedom, as defined above, is just one. For simplicity, we start considering a superfluid in its ground state. With no interaction, all the boson particles
occupy the zero momentum state. A steady uniform flow is obtained by moving the whole system at velocity $ \mathbf{v} $, in which case each particle
acquires a momentum $ \mathbf{k} = m \mathbf{v} $ with a wave function $ e^{\imath \mathbf{k} \mathbf{r}} $. The total wave function is then \cite{NoPi}
\beq
\Psi_{0k}({\mathbf{r_i}}) \,=\, \prod_{i} e^{\imath \mathbf{k} \mathbf{r_i}} \,=\, e^{\imath \mathbf{k} \sum_i \mathbf{r_i}} \,=\, e^{\imath \mathbf{K} \mathbf{R}} 
\label{eq:flow}
\eeq
\noindent where $ \mathbf{K} = N \mathbf{k} $, with $N$ the number of particles, and $\mathbf{R}$ the center of mass coordinate. This result can be generalized to the interacting case \cite{NoPi}, since it is a consequence of the uniform
translation
\beq
\tilde{\Psi}_{0k}({\mathbf{r_i}}) \,=\, e^{\imath \mathbf{K} \mathbf{R}} \tilde{\Psi}_{0}
\label{eq:flow1}
\eeq
\noindent where $ \tilde{\Psi}_{0} $ is the exact interacting ground state. One can see that also in this case the number of degrees of freedom is the one of a single particle. This can be seen also by considering the macroscopic wave function of
the superfluid, that can be defined in general as \cite{NoPi} 
\beq
\phi(\mathbf{r}) \,=\, < N-1 | \psi(\mathbf{r}) | N >
\label{eq:macwf}
\eeq
\noindent where $\psi$ is the annihilation operator for the bosons and the matrix element is between condensate states with
$N$ and $N-1$ particles. This matrix element receives the main contribution from the condensate particles.
\par 
For a superconductor one can repeat the procedure for the electron Cooper pairs in place of the boson particles.
The ground state can be written as the product of a many two particle states
\beq
\Psi_0(\mathbf{r_i}) \,=\, \mathcal{A} \prod_{<ij>} w(\mathbf{r_i} - \mathbf{r_j})
\label{eq:supgs}
\eeq 
\noindent where the product is extended to all distinct electron pairs and $ \mathcal{A} $ indicates anti-symmetrization.
This means that all the pairs are condensed in the center of mass zero momentum. Moving the system at velocity $\mathbf{v}$,
each pair gets a momentum $ \mathbf{p} = 2m \mathbf{v} $ and their wave function is multiplied by $ e^{\imath  \mathbf{p} \mathbf{r}} $. Accordingly, the total wave function is multiplied by $ e^{\imath  \mathbf{P} \mathbf{R}} $, with
$ \mathbf{P} = N \mathbf{p}/2 $. This result extends readily to the interacting ground state.
The macroscopic wave function is in this case the amplitude \cite{Schri} 
\beq
\phi(\mathbf{r}) \,=\, < N-2 | a_{\downarrow}(\mathbf{r}) a_{\uparrow}(\mathbf{r}) | N >    
\label{eq:superc}
\eeq
\noindent where $ a_(\mathbf{r}) $ are annihilation operators for electrons with down and up spin, respectively.
One can then conclude that the macroscopic current can be in a superposition of state, without reduction, for an arbitrary number of electrons. As stressed in ref. \cite{SQUID1}, the internal superconducting degrees of freedom remain "frozen" during the dynamical evolution of the system and do not participate to the state superposition.
\par
Similar considerations apply to the case of macroscopic superposition of mechanical states \cite{mac_phonon_mode,drums1,drums2}.
One can picture the macroscopic states as a condensation of phonons in a single mode, and in fact the description of the system excitation is in terms of a single wave function.  
\par
The general conclusion, on the basis of the new criterion, is that so far no transition from quantum to classical regime has been observed in the experiments because the number of degrees of freedom has been always one or few units.
\section{A possible scheme for the classical limit. \label{sec:general}}
The assumption that the transition from quantum to classical physics 
occurs at a critical number of degrees of freedom appears to be compatible with the experimental observation. If this criterion
is correct, it would introduce a new paradigm, since the number of degrees of freedom can be much smaller than the number of particles. The latter can be arbitrarily large, while the number of degrees of freedom can be quite limited.\par 
The transition from QM to classical mechanics must occur in a strong quantum measurement. In fact the response of e.g. a detector must be describable in classical terms, i.e. at macroscopic level. The crucial step along this transition is the wave function
reduction. Typically, at the beginning of the measurement the apparatus, together with the detected system, follow the linear evolution of the Schrodinger equation, and in general it will be in a superposition of the possible responses. The latter must undergo a process of amplification toward the appearance of a macroscopic signal, and therefore the number of degrees of freedom
must increase with time. If the critical value is reached, the reduction takes place and only one of the responses is selected,
following Born's rule.
The disappearance of the superposition is one of the main prerequisites for the occurrence of the classical limit, which eventually will be established by further amplification, when the size of the selected response will be large enough that it can be observed without perturbation. This picture of the measurement process emerges automatically
in the model of ref. \cite{IJQF}, where an extension of standard QM is proposed. In the model the reduction process in a measurement finds a natural explanation and at the same time it suggests how the reduction occurs when the number of degrees of freedom exceed a given threshold. Unfortunately this limiting value cannot be fixed by the model, but it is very likely to exist. \par 
It has to be stressed that in this picture of the measurement we have assumed that the evolution occurs within the phase space of the possible responses of the apparatus. This defines a "perfect" detector, i.e. one which can be excited only in a definite set of states by the detected system. This identifies uniquely the representation which describes the quantum state of the apparatus. As an example we can consider again a scintillator, where the responses are the excitations of its atoms or molecules, located at different positions within the material. In fact an incoming particle interacts with the scintillator with a finite range interaction. Since the scintillator has no extended collective excitations the particle can only excite locally the atoms or molecules of the detector. If the particle is described by an extended wave packet, it can virtually excite the scintillator in different positions, i.e. the state of the detector is indeed a superposition of the possible responses.   \par   
Another interesting example is the detection of an alpha particle emitted by a radioactive nucleus in a Wilson chamber. Here one observes the linear trajectory traced by the formation of water droplets along the path of the particle. The particle in fact ionizes the water molecules of a supersaturated steam vapor, which become the nucleation seeds of the droplets. One then notices that, as it is well known, the alpha particle is actually emitted in the s-wave, since it is the only way to avoid the centrifugal barrier. This means that the emitted particle is described originally by an isotropic wave packet, i.e. all the possible directions have equal probabilities. Each direction corresponds to a possible response of the detector, i.e. the trace of a linear trajectory. 
As the size of the virtual droplets reach a large enough size, reduction takes place and one direction is selected, which gives rise to the observed linear trajectory. Also in this case the interaction is short range and acts locally, which makes possible the distinction among the different directions and different responses. Here the number of degrees of freedom of a nucleating droplet coincides with the number of particles, since the water molecules can move freely inside the droplet. The droplets are of course classical objects, since they can be observed with naked eyes.\par 
Besides measurement, in general
the classical limit of QM, both at observational level and at formal level, has many facets and it requires a careful analysis of the physical situation that can be expected in this delicate limit. First of all it is obvious that for a massive physical object the internal degrees of freedom must follow QM, since they correspond to e.g. interacting atoms or molecules, irrespective of the size of the system. Even the random motion of atoms or molecules in a dilute gas must follow QM, since they are microscopic elementary objects. In considering the classical limit it is therefore mandatory to select those variables that are used to describe a massive object at classical level, e.g. the center of mass position and the orientation. These are the "macroscopic variables" that can undergo a possible transition from quantum to classical behavior, since the internal degrees of freedom must be always quantal. As an example of other macroscopic variables one can mention the deformation or the vibrations of the massive object.\par 
One crucial question of the QM formalism is if and how the wave function reduction takes place besides the measurement process.
In general the definition of the classical limit is a little vague, so that, for the later discussion, we define the requirements that a physical system must fulfill to be considered "classical" or to have a "classical behavior",
exactly or valid to all purposes.
\par\noindent
i) {\it The superposition between quantum states of macroscopic variables is absent or impossible.} 
\par\noindent 
ii) {\it The relevant macroscopic variables must be sharp and follow classical mechanics.} 
\par\noindent
iii) {\it Any measurement on the system must have negligible effects on the system.} 
\par 
Requirement i) is especially relevant for the paradoxes like the "Schrodinger cat" paradox. In this specific case along the two paths that lead, one to the "cat alive", the other to the "cat dead", the number of degrees of freedom increases, since the materials that allow the dynamic to evolve are necessarily excited. At the critical value the reduction occurs. The final stage
obviously fulfill the other two requirements. The process can be viewed as a sort of measurement process, where the cat and the whole apparatus are mainly the detector. The paradox shows that ordinary QM cannot be extended arbitrarily to macroscopic objects.\par 
Let us consider a physical object, eventually subject to an external force. If the system is close enough to its ground state,
the only relevant macroscopic variables are the center of mass position and momentum, apart its possible orientation.
It is well known that the evolution of a quantum wave packet which is close to a classical state can remain close to the corresponding classical trajectory only for a limited lapse of time, usually referred as 'Ehrenfest time' $t_E $ \cite{Ehren}. Above this time the modulus square of the wave packet spreads out of the classical trajectory and no strict classical limit appears to be possible. In a region of unstable classical trajectories this time is expected \cite{Berman} to be not longer than a limiting time which increases logarithmically as $ ln(S/\hbar) $ in the limit $ \hbar \rightarrow 0 $. Here $S$ is a typical classical action of the system. Therefore, in this case, even for macroscopic system, the so called "correspondence principle" appears to be untenable even for relatively short time. This appears a solid result, since any perturbation of the object cannot make the wave packet to "disappear" in favor of a classical behavior. It is compatible with the experimental
observation of ref. \cite{large}, discussed in sec. \ref{sec:exp}, where one observes the persistence of quantum behavior for the stretched wave packet of the center of mass motion of a cold atom cloud.  
 If we move away from the ground sates the number of degrees of freedom, as defined above, will increase. Above the critical value reduction takes place. In which representation this has to occur will depend in general on the physical conditions of the system, as will be discussed below. In the considered model reduction is produced by a stochastic interaction among the components and therefore one has first to establish how the stochastic equation for this process depends on the representation.
The stochastic part of the equation of motion is non linear and therefore the usual linear transformation from one representation to another cannot be anymore applied. However one can notice that the form of the stochastic equation is a direct consequence of the conservation of the standard norm \cite{IJQF}, which is a necessary requirement if the notion of quantum state has to be taken, independently from its probabilistic interpretation. Therefore, the form of the equation must be the same in any representation.\par
Let us consider the CM degrees of freedom, which is characterized by a wave packet. Keeping the same internal excited state, before reduction the different components correspond to the different possible position of the CM. Each one of these extra degrees of freedom has to be taken with respect to the CM, and therefore the physical situation is equivalent to the case of the scintillator discussed above, where each component is located in different position in ordinary space. However in this case there is no detector that fix the representation, and at the critical value of the degrees of freedom the reduction must occur also in momentum space. Then the uncertainty principle demands that the reduction process will ultimately select a minimum wave packet \cite{Messiah}. As sown below, the wave packet cannot expand, contrary to the standard QM, since reduction will instantaneously keep its size, but it will follow the Ehrenfest classical equation of motion with no spread. This is clearly a classical behavior, which become exact in the short wavelength limit, where the quantum fluctuations can be safely neglected.   
One can say that the internal excitation of the system hinder the quantum coherence in the CM motion. To see that, let us remind that the minimum wave packet has a gaussian shape and for a free motion
its CM position $X$ and width $\chi$ evolve after a time $\Delta t$ according to the equations \cite{Messiah} 
\begin{eqnarray}
X(\Delta t) &\,=\,& X_0 \,+\, \frac{p_0}{m} \Delta t \\ \nonumber
\ &\ & \\ \nonumber
\chi{\Delta t}) &\,=\,&  \big[ \chi_0^2 \,+\, \frac{\xi^2}{m^2} {\Delta t}^2 \big]^{\frac{1}{2}} \,\sim\,
\chi_0 \,+\, \frac{\xi^2}{2 m^2 }\, {\Delta t}^2/\chi_0^2 \,+\, \cdots                                                                                                                                                                                                                                                                                                                                                                                                                                                                                                                                                                                                                                                                                                                                                                                                                                                                                                                                                                                                                                                                                                                                                                                                                                                                                                                                                                                                                                                                                                                                                                                                                                                                                                                                                                                                                                                                                                                                                                                                                                                                                                                                                                                                                                                                                                                                                                                                                                                                                                                                                                                                                                                                                                                                                                                                                                                                                                                                                                                                                                                                                                                                                                                                                                                                                                                    
\label{eq:spread}
\end{eqnarray}
\noindent where $p_0$ and $\xi$ are the average and spread in momentum, both independent of time. Therefore its width increases indeed indefinitely. This means also that after a time $\Delta t$ the initial wave packet is centered at $X(\Delta t)$ but it is not any more a minimum one and therefore it is not any more gaussian. The deviation from a minimum one is of order ${\Delta t}^2$,
i.e. the wave packet  can be written
\beq
w_{\Delta t}(x) \,=\, w_0(x+X(\Delta t)) \,+\, \Delta t^2 w_1(x) \,+\, \cdots
\label{eq:corr}   
\eeq
\noindent where $w_0$ is the initial minimum wave packet and $w_1$ a residual function that eventually can be expanded in minimum wave packets. If at time $\Delta t$ a reduction process takes place, according to Born's rule the first term will be selected with probability $1$, with negligible ${\Delta t}^4$ corrections. The resulting wave packet will be the original minimum wave packet shifted at $X(\Delta t)$, and no spread occurs. In the limit of vanishing $\Delta t$ the minimal wave packet will follow the classical trajectory exactly and unaffected. This result can be extended to the presence of a potential,
provided it is sufficiently smooth. Despite QM is still active, the inclusion of wave function reduction enforces the appearance of the classical behavior in the dynamics of a single object moving in a potential.
\section{Conclusion. \label{sec:conc}}
The laws of QM is quite different from the ones established at the classical level. It is therefore necessary that a transition from QM to classical physics occurs as we move from the microscopic to the macroscopic worlds. This transition could be smooth or abrupt, and it could be exact or approximate. The latter possibility has to be excluded because the extension of the superposition principle to macroscopic level produces well defined paradoxes. The most known is the so called "Schrodinger cat paradox", where a cat (or any living being of macroscopic size) is found to be in a quantum superposition of dead and alive states. 
Unfortunately this transition is absent in the formalism of ordinary QM. In fact the suppression of the superposition principle is not viable in the standard formulation of QM, because this necessarily implies the invalidation of the Schrodinger equation.
\par 
It is commonly believed that the transition should appear at some large enough value of the number of particles of a given physical system. Many experiments, discussed in section \ref{sec:exp}, have however shown that the superposition of quantum states still persists even for system that contains a macroscopic number of particles. In this paper we suggest that the relevant parameter for the occurrence of the transition is not the number of particles, but rather the number of degrees of freedom, properly defined. In general this parameter is smaller, or even much smaller, than the number of particles. This assumption is in agreement with one of the result in the model of ref. \cite{IJQF}, which completes the standard QM, and it allows the inclusion of the additional process of wave function reduction, which violates the Schrodinger equation. We have seen that in all the considered experiments the number of degrees of freedom is one or very small, which would explain why standard QM is still valid even for a macroscopic size of the system.\par 
We have then considered the motion of a macroscopic object and shown that the presence of the wave function reduction, above a critical value for degrees of freedom, hinders the QM spread of the CM wave packet, which entails an effective classical behavior
of its dynamics in free space.\par 
To check the validity of this new paradigm would be necessary to devise experiments with an increasing large number of degrees of freedom. This appears quite challenging, and in any case well beyond the scope of the present paper.

\end{document}